\documentclass[showpacs,twocolumn,amsmath,amssymb,superscriptaddress]{revtex4-1}

\usepackage{color}
\usepackage{graphicx}
\usepackage{amssymb}
\usepackage{amsmath}
\usepackage{mathrsfs}
\newcommand{\ud}{\mathrm{d}}
\newcommand{\diag}{\mathrm{diag}}

\newcommand{\ket}[1]{|#1\rangle}

\newcommand{\de}{\partial}

\newcommand{\eps}{\epsilon}

\usepackage[breaklinks=true]{hyperref}
\graphicspath{ {./figures/} {./}}

\begin{document}

\title{Suppression and splitting of modulational instability sidebands in periodically tapered optical fibers due to fourth-order dispersion}

\author{Andrea Armaroli}
\email{andrea.armaroli@mpl.mpg.de}
\affiliation{Max Planck Research Group `Nonlinear Photonic Nanostructures' \\Max Planck Institute for the Science of Light, G{\"u}nther-Scharowsky-Str.~1/Bau 24
91058 Erlangen, Germany}
\author{Fabio Biancalana}
\affiliation{Max Planck Research Group `Nonlinear Photonic Nanostructures' \\Max Planck Institute for the Science of Light, G{\"u}nther-Scharowsky-Str.~1/Bau 24
91058 Erlangen, Germany}
\affiliation{School of Engineering and Physical Sciences, Heriot-Watt University, EH14 4AS Edinburgh, United Kingdom}


\begin{abstract}
We study the modulational instability induced by periodic variations of group-velocity dispersion  in the proximity of the zero dispersion point. Multiple instability peaks originating from parametric resonance coexist with the conventional modulation instability due to fourth order dispersion, which in turn is suppressed by the  oscillations of dispersion. Moreover isolated unstable regions  appear in the space of parameters due to imperfect phase matching. This confirms the dramatic effect of periodic tapering in  the control and shaping of MI sidebands in optical fibers.
\end{abstract}

\maketitle


In classical mechanics parametric resonance (PR) is a well-known instability phenomenon which occurs in systems the parameters of which are varied periodically during evolution \cite{ArnoldCM,LandauCM}. Consider, for example, a harmonic oscillator the frequency of which is forced to vary in time. There exist regions in the parameter space (amplitude of variation vs.~natural frequency), known as {\em resonance tongues}, where the oscillator is destabilized. 

It is natural that such a general phenomenon was associated to the equally important instability process that is ubiquitous in infinite dimensional dynamical systems: modulation instability (MI)
which in nonlinear optics \cite{Karpman1967} manifests itself as pairs of sidebands  exponentially growing  on top of a plane wave initial condition, by virtue of the interplay between the cubic Kerr nonlinearity and the group velocity dispersion (GVD). 

The link between PR and MI has been established  in relation to the periodic re-amplification of signals in long-haul telecommunication optical fiber cables \cite{Matera1993}. This was based on a nonlinear Schr{\"o}dinger equation (NLS) where the coefficient of the nonlinear term is varied along the propagation direction. Importantly, this peculiar type of MI occurs in both normal and anomalous GVD. 
Moreover, in long-haul fibers, dispersion management is a commonly used technique which introduces periodic modulation of fiber characteristics. The possibility of instability phenomena disrupting adjacent communication channels has been thoroughly analyzed, see e.g.~\cite{Smith1996,Bronski1996c,Kumar2003,Ambomo2008}.  Similar phenomena are also found in other branches of physics, such as Bose-Einstein condensates with time-varying external fields \cite{Abdullaev2013,Ramaswamy2014}.

The occurrence of these effects in micro-structured fibers, which permit to observe them on a shorter scale and explore a wide range of input power and wavelength, have been recently reported, see \cite{Droques2012,Droques2013}, where a  photonic-crystal fiber (PCF, \cite{RussellScience2003}) of varying diameter is used. By operating near the zero-dispersion point (ZDP) \cite{Droques2013a}, the main contribution comes from the variation of the GVD, from which the name dispersion-oscillating fibers (DOFs);
 partial coherence and the effects of large GVD swings \cite{Finot2013,Finot2014} were considered. 

The conventional  explanation is in term of a long grating assisted quasi-phase matching (QPM) process \cite{Matera1993, Kumar2003, Ambomo2008, Droques2012,Droques2013}, but it was verified in \cite{Armaroli2012} that this approximation is inaccurate if the period of parameter variation is comparable with the length scale at which the nonlinear processes occur. In \cite{Armaroli2012} it was proved that an accurate description must be based on the Floquet theory \cite{LandauCM,ArnoldGMODE} and the use of regular perturbation techniques, such as the method of averaging \cite{VerhulstBook2010}. Thanks to those techniques we found new instability sidebands of parametric origin in highly-birefringent optical fibers \cite{Armaroli2013}.

In this work we thoroughly study the effect of higher-order dispersion  of the fourth order (FOD). 
Particularly a large dispersion oscillation suppresses the conventional MI mechanism; moreover we observe twisted resonance tongues, which result in multiple peaks; finally we discover unstable regions corresponding to imperfect phase-matching. By combining numerical and analytical techniques, we definitely improve the understanding of this phenomenon and show the variety of instability mechanisms which occur in these systems, which were not completely quantified in \cite{Droques2013a}. 



We consider the adimensional generalized NLS including FOD,
\begin{equation} \label{eq:NLS1}
i\de_{z}A- \frac{1}{2}\beta_2(z)\de_{t}^{2}A + \frac{1}{24}\beta_4 \de_{t}^{4}A+\gamma(z)|A|^2A=0.
\end{equation}
$\beta_2$, $\beta_4$  and $\gamma$ are normalized dispersion (of second and fourth order) and nonlinear coefficients, $\beta_2(z)\equiv\overline\beta_2(z)/\overline\beta_2^0$, $\beta_4=\overline\beta_4/(T_0^2\overline\beta_2^0)$ and $\gamma(z) \equiv \overline\gamma(z)/\overline\gamma^0$, where $\overline\beta_2(z)$,  $\overline\beta_4$ and $\overline\gamma(z)$ are the GVD, FOD and nonlinear coefficients, respectively, and the $0$ superscript denotes their mean values. $\beta_2$ and $\gamma$  are assumed to be periodic functions of $z$, while $\beta_4$ is supposed to give a constant small contribution. 
Moreover $z\equiv Z/Z_{nl}$ is the dimensionless distance in units of the nonlinear length $Z_{nl}\equiv(\gamma^0 P_{t})^{-1}$, and $t\equiv(T-Z/v_g^0)/T_{s}$ is the dimensionless retarded time in units of  $T_{s}\equiv \sqrt{Z_{nl} |\beta_2^0|}$, and $v_g^0$ is the mean group velocity. 
Finally $P_t$ is the total input power injected in the mode, and $A$ is the dimensionless modal intensity scaled by $\sqrt{P_{t}}$.

We look for  a steady state solution of \eqref{eq:NLS1} in the form $A=\sqrt{P_0}\exp{(i \Phi(z))}$: it can be verified that $\Phi(z) = P_0\int_{-\infty}^{z}{n(z')\ud z'}$. We then perturb this steady state by adding a small complex time dependent contribution $a(z,t)$, i.e.~$A(z,t)=\left(\sqrt{P_0}+\varepsilon a(z,t)\right)\exp{(i \Phi(z))}$, with $\eps\ll 1$. Inserting this \emph{Ansatz} in Eq.~\eqref{eq:NLS1} and taking only the terms which are first order in $\eps$, one finds that  $a$ obeys the following equation:
\begin{equation}
	i\partial_z a -\frac{1}{2}\beta_2(z)\partial^{2}_{t}a +\frac{1}{24}\beta_4\partial^{4}_{t}a +\gamma(z)P_0(a+a^*) = 0.
\label{eq:NLS1lin}
\end{equation}
Finally we expand $a$ in \eqref{eq:NLS1lin} as a sideband pair, 
$
a(z,t) = a_A(z) e^{-i \omega t} + a_S(z) e^{i \omega t}
$
to obtain the following $z$-dependent Schr\"odinger equation
\begin{equation}
	i\dot{\ket{\psi}} = H_{\rm s}(z) \ket{\psi},
	\label{eq:NLS1SAS}
\end{equation}
where
\[H_{\rm s}(z) \equiv \left(-\beta_2(z)\frac{\omega^2}{2}-\beta_4 \frac{\omega^4}{24}-\gamma(z) P_0\right)\hat{\sigma}_z - i \gamma(z)P_0 \hat{\sigma}_y,\]
 the dot denotes $z$-derivative, 
$ \ket{\psi} \equiv (a_A,a_S^*)^T$ and
 $\hat{\sigma}_i$ are the Pauli matrices. 
We assume for the parameters of Eq.~\ref{eq:NLS1} a simple harmonic dependence
\begin{equation}
\begin{gathered}
	\beta_2(z) = \beta_0 + \tilde \beta(z) = \beta_0+h \beta_1\cos{\Lambda z}, \\
	\gamma(z) = \gamma_0+\tilde \gamma(z) = \gamma_0+h \gamma_1\cos{\Lambda z},
\end{gathered}
	\label{eq:dispnlcos}
\end{equation}
where generally $\beta_0=\pm1$ for normal (anomalous) GVD and $\gamma_0=1$; $\Lambda$ is the normalized spatial angular frequency for the parameter oscillations. The forcing amplitude is controlled by the parameter $h$. We will observe two regimes: if $h\ll 1$ we can analytically estimate the unstable regions, by means of the method of averaging; if $h\sim 1$ QPM arguments are more expedient. 

Let us first transform it in phase-quadrature variables, by applying a  rotation operator as in \cite{Armaroli2012} and
thus study the system
\begin{equation}
i\dot{\ket{\phi}} = H_{\rm pq}(z) \ket{\phi}
\label{eq:NLS1PQ2}
\end{equation}
with
\begin{equation}
	H_{\rm pq} = 
	\begin{pmatrix}
0&c_1(z)\\c_2(z)&0
\end{pmatrix}
\label{eq:NLS1PQ}
\end{equation}
with $c_1(z) = -\frac{1}{2}\beta_2(z)\omega^2-\frac{1}{24}\beta_4\omega^4$ and $c_2(z) = c_1(z)-2\gamma(z)P_0$.

By replacing \eqref{eq:dispnlcos} in these definitions we can naturally split the Hamiltonian matrix $H(z)$ into average and oscillating parts, i.e. $H(z)\equiv H_0 + h\tilde{H}(z)$.

In the limit of vanishing perturbation $h\to 0$, we have a simple harmonic oscillator written for any of the components of $\ket\phi$. It is widely known that, in this limit, parametric resonance occurs if the natural frequency of the oscillator is a multiple of half the forcing frequency, see \cite{LandauCM,Armaroli2012,Armaroli2013,Droques2013a}. Thus, by setting $c_{1,2}(z)=c_{1,2}^0+\tilde{c}_{1,2}(z)$,  the natural frequency of the unforced oscillator is simply $
\lambda_0=\sqrt{c_1^0c_2^0}$ and the resonance condition is
\begin{equation}
c_1^0c_2^0 = \left[ \frac{m \Lambda}{2}\right]^2,
\label{eq:PRcond}
\end{equation}
with $m=1,\,2,\,3,\ldots$ the PR order.
This is an algebraic equation and its roots correspond to the resonant detuning $\omega_m$.
In general one or three solutions exist according to the values of $\beta_4$.

\begin{figure}
	\centering
		\includegraphics[width=0.40\textwidth]{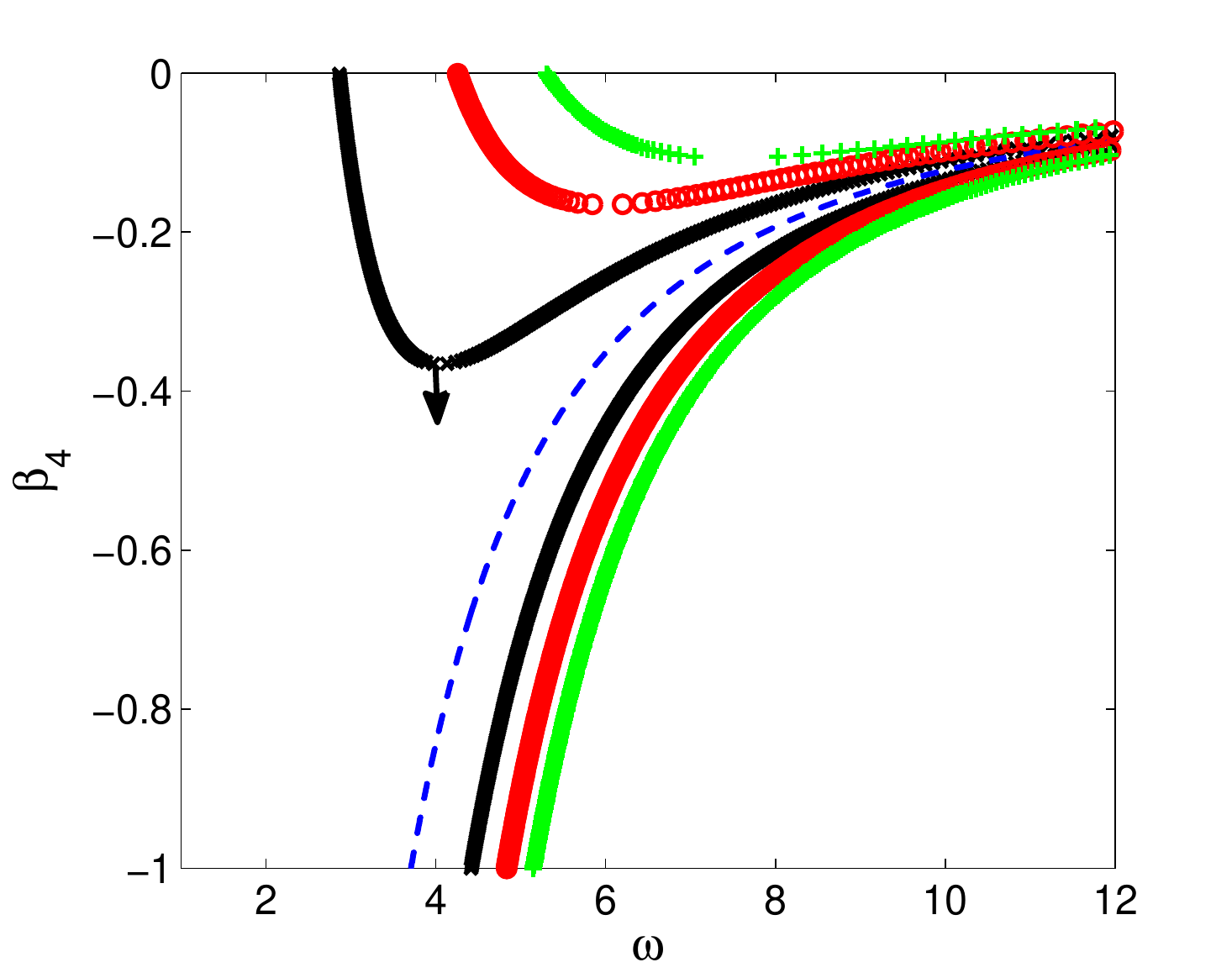}
	\caption{Position of instability peaks as a function of FOD $\beta_4$ for first (black crosses), second (red circles) and third (green plus) order parametric resonance. For comparison the corresponding conventional MI (blue dashed line) is also reported. The arrow shows the approximate region where the instability occurs despite the resonance condition (or phase matching) cannot be perfectly achieved.}
	\label{fig:PRfrequencies}
\end{figure}

In Fig.~\ref{fig:PRfrequencies} we show the PR detuning ($\Lambda=10$) for the $m=1,2,3$ and include also the conventional MI, which occurs in normal dispersion at $\omega_0^2 = - \frac{6\beta_2^0}{\beta_4}+\sqrt{\left(\frac{6\beta_2^0}{\beta_4}\right)^2-\frac{24\gamma_0P_0}{\beta_4}}$. 
This sideband occurs between one large detuning and  (possibly) a pair of small detuning PR sidebands.
In fact, at $\overline\beta_4 = -\frac{3}{(m \Lambda)^2}\left[2+\sqrt{4 + (m \Lambda)^2}\right]=-0.366$, two PR collide and only the high frequency solution appears to survive. Below we show this is only partially true, because the instability is strongly affected by the value of $h$.


The sideband gain and bandwidth can  be easily obtained for $m=1$ by means of the method of averaging \cite{VerhulstBook2010}. 
This in turn is equivalent to transforming the system of Eq.~\eqref{eq:NLS1PQ} to the interaction picture of quantum mechanics with a perturbed eigenvalue pair, i.e.~we assume the eigenvalues of $H_0$ ($\pm\lambda_0$) are perturbed $\lambda_0 \to \lambda_0+\varepsilon$ and define $H_0' = V\Lambda' V^{-1}$, with $\Lambda' = \diag\left\{\lambda_0+\varepsilon,-\lambda_0-\varepsilon\right\}$ and $V$ the matrix the column of which are the eigenvectors of $H_0$.
The evolution of the slow variables  $\ket{\phi}_I=e^{i H_{0}' z}\ket{\phi}$ is governed by
\begin{equation}
	i\ket{\dot\phi}_I = h  H_I(z) \ket{\phi}_I, \text{ with }
	H_I = e^{i H_{0}' z} \tilde{H}(z) e^{-i H_{0}' z}. 
	\label{eq:NLSPQ2}
\end{equation}
The  averaging process is used to eliminate the remaining oscillating terms by solving Eq.~\eqref{eq:NLSPQ2} with the $z$ averaged (over one period $2\pi/\Lambda$) Hamiltonian  $\langle H_I(z)\rangle$.

By applying this method we obtain again the resonance condition in Eq~\eqref{eq:PRcond}.  
Moreover, around the PR detuning we can estimate the peak gain (setting $\varepsilon=0$) to be $g=\kappa/4$ and the instability margins (imposing vanishing gain) occurring for $\lambda_0'=\lambda_0\pm\kappa/4$, where 
\[
\kappa = \lambda_0\left(\frac{\tilde{c}_2}{c_2^0}-\frac{\tilde{c}_1}{c_1^0}\right)
\]
and we extract the detuning at the margin by finding the roots $\omega_1\pm\Delta\!\omega_1$ of the algebraic equation $	\lambda_0'=\Lambda/2$. This constitutes our first result.

Then we compare these estimates with the exact solution obtained by solving numerically the Floquet problem associated to Eq.~\eqref{eq:NLS1SAS} or \eqref{eq:NLS1PQ} and obtain the regions of the $(\omega,h)$ plane, where the system is unstable, the instability tongues. We set $\Lambda=10$ and $\gamma_1=0$, $P_0 = 1$, thus vary the dispersion only by controlling the parameter $h$. 

\begin{figure}
	\centering
		\includegraphics[width=0.40\textwidth]{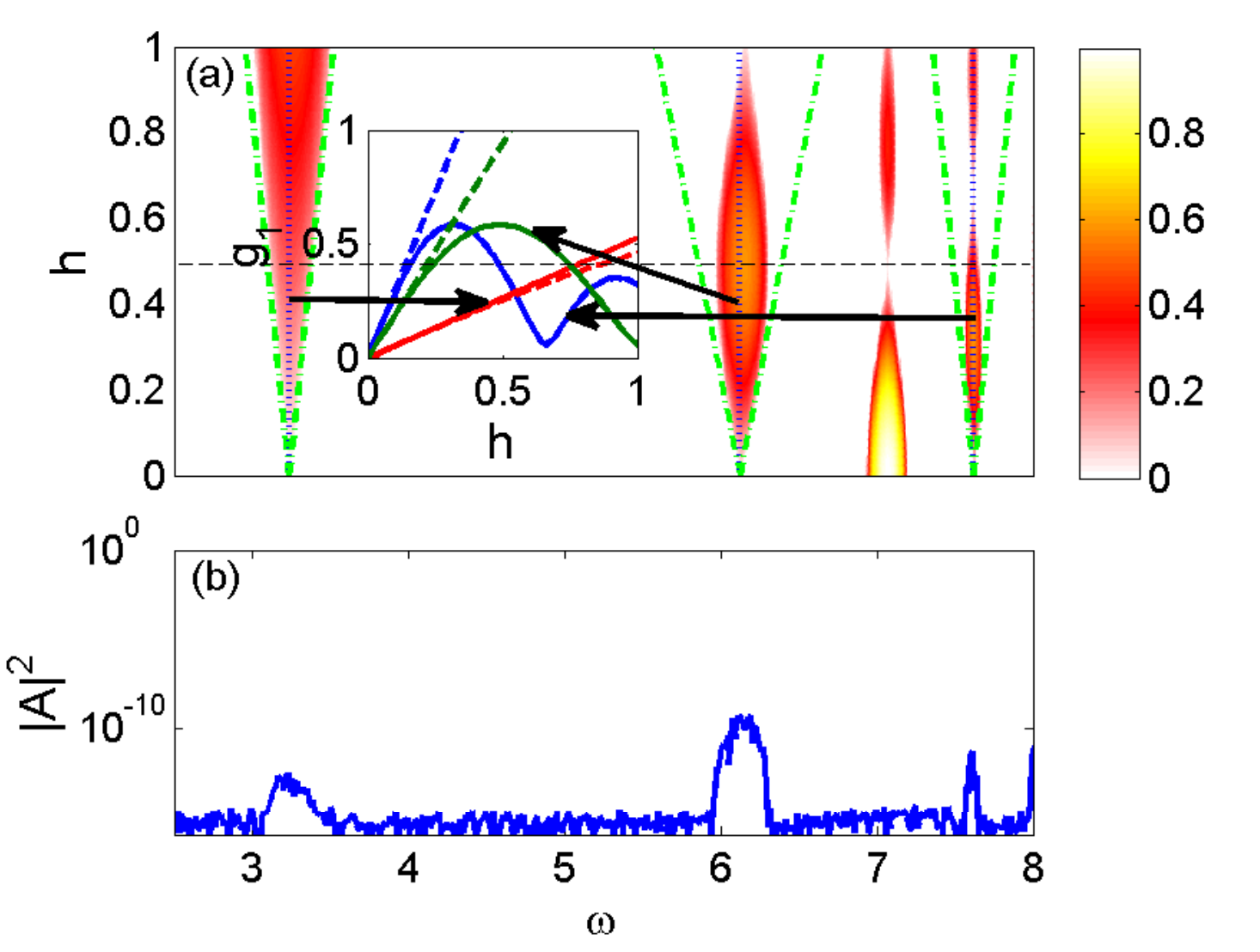}
		\caption{(Color online) (a) Resonance tongues for 1\textsuperscript{st} order PR, with $\Lambda=10$ and $\beta_4=-0.25$. The color scale corresponds to the instability gain The dotted blue lines denote the predicted positions of PR peaks, while the green dash-dotted lines to the predicted instability margins. The corresponding maximum gain is showed in the inset (solid lines) as a function of the perturbation strength $h$ and is compared with the analytical predictions (dashed lines). The arrows connect each gain curve with the corresponding resonance tongue. (b) Output of split-step Fourier simulation of Eq.~\eqref{eq:NLS1} at $z\approx14$, for $h = 0.5$ (corresponding to the dashed horizontal line in panel (a). }
	\label{fig:PR0p25}
\end{figure}
In Fig.~\ref{fig:PR0p25} we show a typical example of the shape of unstable regions for small $\beta_4=-0.25$ (far from the ZDP), where three PR sidebands occurring for $m=1$. We include in the picture the analytical estimates and we observe that at small $h$ the agreement is almost perfect. 
\begin{figure}
	\centering
		\includegraphics[width=0.40\textwidth]{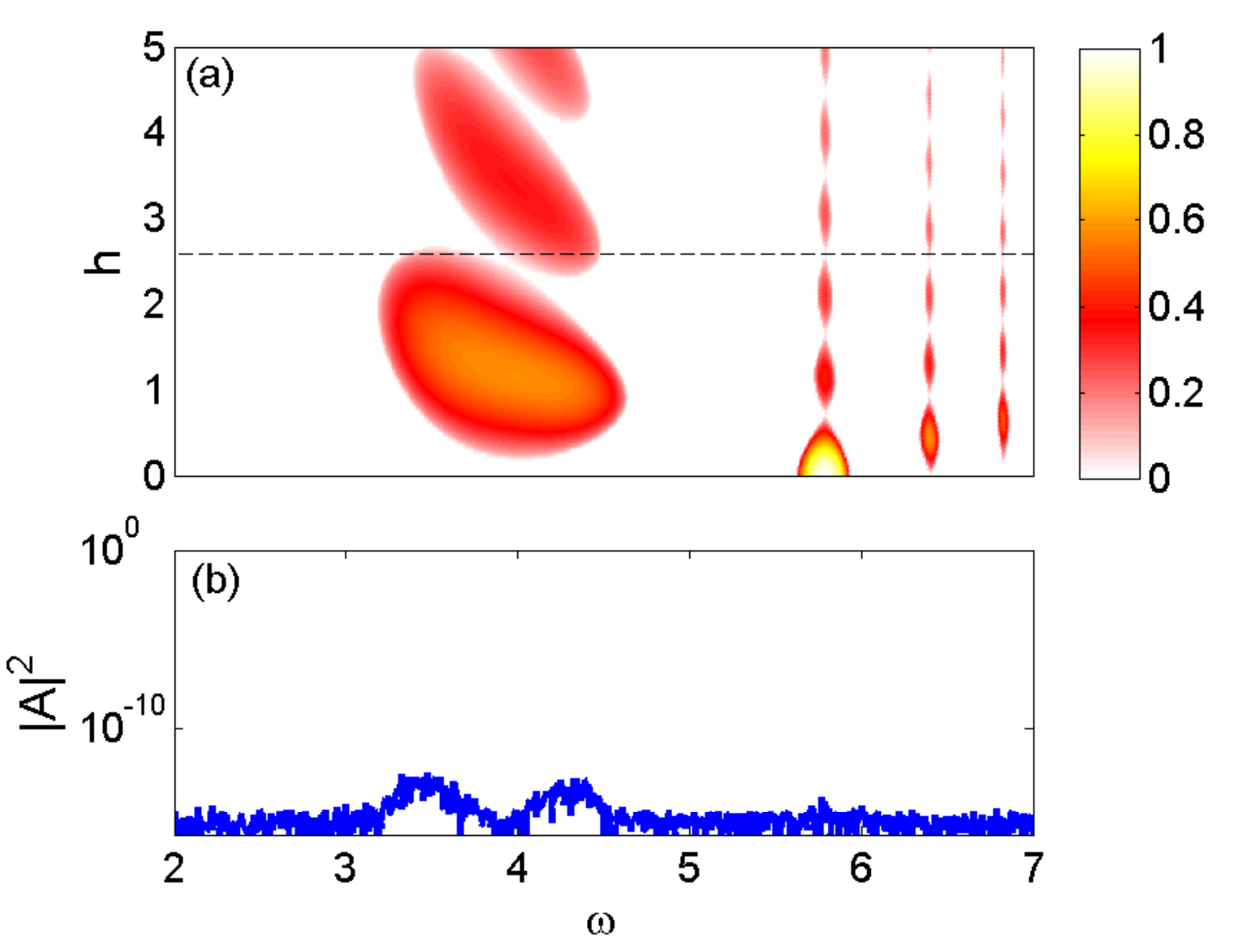}
		\caption{(Color online) Same as in Fig.~\ref{fig:PR0p25}, but with $\beta_4=-0.38$ and a larger $h $ range. Despite phase matching of low detuning sidebands is not achieved, we observe anyway unstable regions for $\omega\approx4$ (a). The simulations in (b) were performed for $h = 2.5$.}
	\label{fig:PR0p38}
\end{figure}
\begin{figure}
	\centering
		\includegraphics[width=0.40\textwidth]{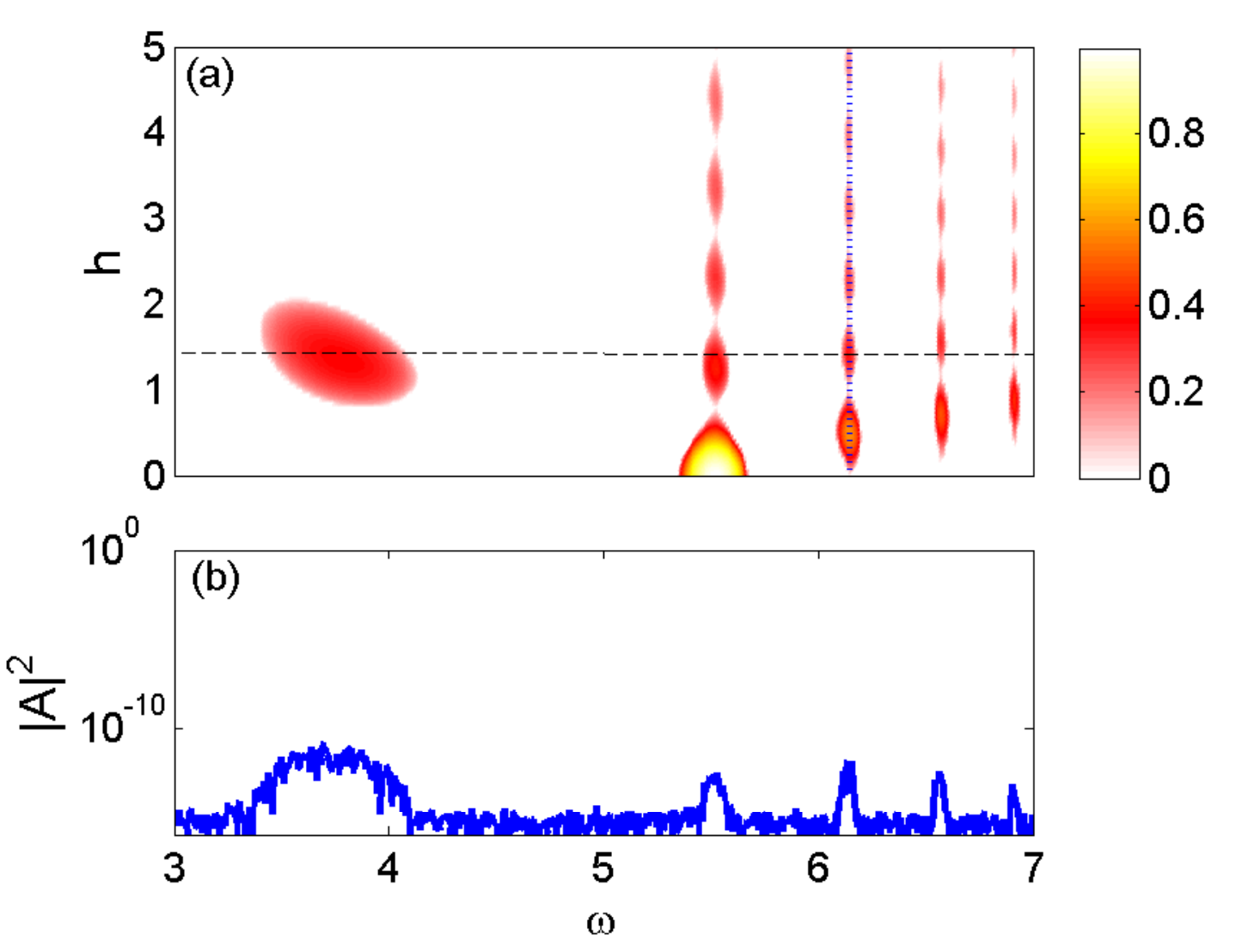}
		\caption{(Color online) Same as in Fig.~\ref{fig:PR0p38}, but with $\beta_4=-0.42$. Despite phase matching of low detuning sidebands is not achieved, we observe anyway an unstable spot  in the parameter space for $\omega\approx4$ and $h\approx1.5$ (a). The simulations in (b) were performed at that value.}
	\label{fig:PR0p42}
\end{figure}
For large values of $h$, the instability tongues (at any order, the conventional MI included) become twisted and it is simpler to resort to QPM arguments.
For a variation of dispersion only, we can write the gain as \cite{Smith1996,Finot2014}
\begin{equation}
g_p^{QPM} = \left[\left(\gamma_0 P_0{J}_p\left(\frac{h\omega^2}{\Lambda}\right)\right)^2-\Delta\!k_{p}^2\right]^{\frac{1}{2}}
\label{eq:QPMgain}
\end{equation}
where 
$\Delta\!k_p = \frac{1}{2}(c_1^0+c_2^0-p\Lambda) = \frac{1}{2}\beta_2^0\omega^2+\frac{1}{24}\beta_4\omega^4+\gamma_0P_0-\frac{p \Lambda}{2}$ 
is the nonlinear phase mismatch
and ${J}_p$ is the Bessel's function of the first kind and order $p$. $\Delta\!k_p=0$ corresponds approximately to the PR condition reported above in Eq.~\eqref{eq:PRcond}, with the caution that $p$ can be a positive or negative integer, while $m>0$ only and $m=|p|$.

The use of Eq.~\eqref{eq:QPMgain} permits to simply understand the evolution of gain at large $h$: if phase matching is achieved, the  gain corresponds to a Bessel's function, as it apparent from the inset of Fig.~\ref{fig:PR0p25}.
 For $p=0$ we obtain that the conventional MI is progressively suppressed for large $h$.  The order $p=-1$ coincides almost perfectly with the high frequency $m=1$ sideband lying beyond the conventional MI peak: it is always phase-matched, oscillates faster along the $h$ axis and exhibits much narrower instability regions.  Instead for $p=1$ as the two low-frequency $m=1$ sidebands in Fig.~\ref{fig:PRfrequencies} merge at $\overline\beta_4$, we cannot obtain exact phase matching anymore, $\Delta\!k_p\neq 0$, but this does \emph{not} imply instability is completely ruled out. In fact, see Fig.~\ref{fig:PR0p38} and \ref{fig:PR0p42} there exist regions of the parameter space where the main instability mechanism stems from those partially phase mismatched PR sideband, provided that $g_p^{QPM}>0$. For $\beta_4$ slightly smaller than $\overline\beta_4$, multiple peaks appear at slightly different detuning, see Fig.~\ref{fig:PR0p38}(a). By decreasing further the FOD, few isolated unstable spots appear, then only a single one, \ref{fig:PR0p42}(a), which finally disappears.

Finally we show in panels (b) of Figs.~\ref{fig:PR0p25}--\ref{fig:PR0p42} the result of NLS simulations for the different scenarios, at different values of $h$. 
We observe a perfect agreement with the prediction of the Floquet theory:   multiple sidebands at each order,  suppression of conventional MI, splitting of sidebands, and peaks appearing at partial phase-matching are completely confirmed.

To conclude, we presented a thorough account of the effects of FOD on MI sidebands appearing in DOFs. We obtained that the conventional MI is suppressed for large amplitude oscillations of dispersion and that for a large enough negative FOD (i.e.~close to the ZDP), the sideband merges and the imperfect phase matching result in unstable regions at small detuning, which, despite the imperfect phase-matching, represent the main instability mechanism in these fibers while large-detuning sidebands have vanishing gain. 

This permits to conceive applications in the controlled emission of photon pairs in periodically tapered PCFs in the proximity of the ZDP.

This research is funded by the German Max Planck Society for the Advancement of Science.
The authors would also like to acknowledge Christophe Finot of the University of Burgundy and Arnaud Mussot and Alexandre Kudlinski of the University of Lille 1 for fruitful discussions.



\begin{thebibliography}{10}
\newcommand{\enquote}[1]{``#1''}

\bibitem{ArnoldCM}
V.~I. Arnold, A.~Weinstein, and K.~Vogtmann, \emph{Mathematical Methods of
  Classical Mechanics (Graduate Texts in Mathematics)} (Springer, 1989).

\bibitem{LandauCM}
L.~D. Landau and E.~Lifshitz, \emph{Mechanics, Third Edition: Volume 1 (Course
  of Theoretical Physics)} (Butterworth-Heinemann, 1976).

\bibitem{Karpman1967}
V.~I. {Karpman} and E.~M. Krushkal, \enquote{{Self-modulation of Nonlinear Plane Waves in
  Dispersive Media},} Pis'ma Zh. Eksp. Teor. Fiz. [JETP Letters] \textbf{6},
  277--279 (1967).

\bibitem{Matera1993}
F.~Matera, A.~Mecozzi, M.~Romagnoli, and M.~Settembre, \enquote{{Sideband
  instability induced by periodic power variation in long-distance fiber
  links.}} Opt. Lett. \textbf{18}, 1499 (1993).

\bibitem{Smith1996}
N.~J. Smith and N.~Doran, \enquote{{Modulational instabilities in fibers with
  periodic dispersion management.}} Opt. Lett. \textbf{21}, 570--2 (1996).

\bibitem{Bronski1996c}
J.~C. Bronski and J.~{Nathan Kutz}, \enquote{{Modulational stability of plane
  waves in nonreturn-to-zero communications systems with dispersion
  management.}} Opt. Lett. \textbf{21}, 937--9 (1996).

\bibitem{Kumar2003}
A.~Kumar, A.~Labruyere, and P.~Tchofo-Dinda, \enquote{{Modulational instability
  in fiber systems with periodic loss compensation and dispersion management},}
  Opt. Comm. \textbf{219}, 221--232 (2003).

\bibitem{Ambomo2008}
S.~Ambomo, C.~M. Ngabireng, and P.~Tchofo-Dinda, \enquote{{Critical behavior
  with dramatic enhancement of modulational instability gain in fiber systems
  with periodic variation dispersion},} J. Opt. Soc. Am. B \textbf{25}, 425--433 (2008).

\bibitem{Abdullaev2013}
F.~K. Abdullaev, M.~\"Ogren, and M.~P. S\o{}rensen, \enquote{Faraday waves in
  quasi-one-dimensional superfluid fermi-bose mixtures,} Phys. Rev. A
  \textbf{87}, 023616 (2013).

\bibitem{Ramaswamy2014}
A.~Bala\ifmmode~\check{z}\else \v{z}\fi{}, R.~Paun, A.~I. Nicolin,
  S.~Balasubramanian, and R.~Ramaswamy, \enquote{Faraday waves in collisionally
  inhomogeneous bose-einstein condensates,} Phys. Rev. A \textbf{89}, 023609
  (2014).

\bibitem{Droques2012}
M.~Droques, A.~Kudlinski, G.~Bouwmans, G.~Martinelli, and A.~Mussot,
  \enquote{{Experimental demonstration of modulation instability in an optical
  fiber with a periodic dispersion landscape.}} Opt. Lett. \textbf{37},
  4832--4 (2012).

\bibitem{Droques2013}
M.~Droques, A.~Kudlinski, G.~Bouwmans, G.~Martinelli, and A.~Mussot,
  \enquote{{Dynamics of the modulation instability spectrum in optical fibers
  with oscillating dispersion},} Phys. Rev. A \textbf{87}, 013813 (2013).

\bibitem{RussellScience2003}
P.~{\relax St.J}. Russell, \enquote{Photonic crystal fibers,} Science
  \textbf{299}, 358--362 (2003).

\bibitem{Droques2013a}
M.~Droques, A.~Kudlinski, G.~Bouwmans, G.~Martinelli, A.~Mussot, A.~Armaroli,
  and F.~Biancalana, \enquote{{Fourth-order dispersion mediated modulation
  instability in dispersion oscillating fibers.}} Opt. Lett. \textbf{38},
  3464--7 (2013).

\bibitem{Finot2013}
C.~Finot, J.~Fatome, A.~Sysoliatin, a.~Kosolapov, and S.~Wabnitz,
  \enquote{{Competing four-wave mixing processes in dispersion oscillating
  telecom fiber.}} Opt. Lett. \textbf{38}, 5361--4 (2013).

\bibitem{Finot2014}
C.~Finot, F.~Feng, Y.~Chembo, and S.~Wabnitz, \enquote{{Gain sideband splitting
  in dispersion oscillating fibers},} \url{http://hal.archives-ouvertes.fr/hal-00981213}.

\bibitem{Armaroli2012}
A.~Armaroli and F.~Biancalana, \enquote{{Tunable modulational instability
  sidebands via parametric resonance in periodically tapered optical fibers},}
  Opt. Express \textbf{20}, 25096 (2012).

\bibitem{ArnoldGMODE}
V.~Arnold, \emph{Geometrical Methods in the Theory of Ordinary Differential
  Equations (Grundlehren der mathematischen Wissenschaften) (v. 250)}
  (Springer, 1988).

\bibitem{VerhulstBook2010}
J.~A. Sanders, F.~Verhulst, and J.~Murdock, \emph{Averaging Methods in
  Nonlinear Dynamical Systems (Applied Mathematical Sciences)} (Springer,
  2010).

\bibitem{Armaroli2013}
A.~Armaroli and F.~Biancalana, \enquote{{Vector modulational instability
  induced by parametric resonance in periodically tapered highly birefringent
  optical fibers},} Phys. Rev. A \textbf{87}, 063848 (2013).

\end{thebibliography}
\end{document}